\newif\ifdraft
\ifx\withoutcomments\undefined
\drafttrue
\fi
\documentclass{llncs}
\usepackage[utf8]{inputenc}
\usepackage{amssymb,amsfonts,amstext,amsmath}
\usepackage[onelanguage,ruled,vlined,linesnumbered,norelsize]{algorithm2e}

\usepackage{multirow}
\usepackage{multicol}

\usepackage{graphics}

\usepackage{tikz}
\usepackage{verbatim}

\newcommand{\mc}[3]{\multicolumn{#1}{#2}{#3}}

\newcommand{\sss}[1]{{\begin{scriptstyle}#1\end{scriptstyle}}}

\newcommand{\coeffs}[1]{\mathsf{\sss{#1}}}

\newcommand{\FF}{\ensuremath{\mathbb F}}
\newcommand{\F}{\ensuremath{\mathbb F}}
\newcommand{\Fp}{\ensuremath{\FF_p}}

\newcommand{\FFpn}[2]{\ensuremath{\FF_{{#1}^{#2}}}}

\newcommand{\Q}{\ensuremath{\mathbb Q}}
\newcommand{\Z}{\ensuremath{\mathbb Z}}
\newcommand{\ZZ}{\ensuremath{\mathbb Z}}

\newcommand{\EE}{\ensuremath{\mathrm E}}

\newcommand{\cO}{\ensuremath{\mathcal O}}

\newcommand{\normoo}[1]{{\|#1\|_\infty}}

\DeclareMathOperator{\Reslt}{Res}
\let\Res\Reslt
\DeclareMathOperator{\Norm}{Norm}

\DeclareMathOperator{\vlog}{vlog}
\DeclareMathOperator{\lc}{lc}

\newcommand{\MNTnames}{Miyaji--Nakabayashi--Takano}
\newcommand{\MOV}{Menezes--Okamoto--Vanstone}
\newcommand{\FR}{Frey--Rück}

\def\cadonfs{\texttt{cado-nfs}}
\ifx\textsubscript\undefined
\def\textsubscript#1{$_{#1}$}
\fi
\def\JLSVi{JLSV\textsubscript{1}}
\def\JLSVii{JLSV\textsubscript{2}}

\usepackage{microtype}
\usepackage{refcount}
\usepackage{hyperref}
\hypersetup{
   pdftitle={Solving discrete logarithms on a 170-bit MNT curve by
     pairing reduction},    
   pdfauthor={Aurore Guillevic, François Morain and Emmanuel Thomé}, 
   pdfsubject={Number Field Sieve}, 
   colorlinks=true,         
   linkcolor=black,         
   citecolor=black,         
   filecolor=black,         
   urlcolor=black,          
}

\title{Solving discrete logarithms on a 170-bit MNT curve 
  by pairing reduction
\thanks{This is the authors' version. The final post-conference proceedings
  version will appear in the LNCS series at Springer 
  \url{http://link.springer.com/bookseries/558}}
}

\author{
Aurore Guillevic \inst{5}\fnmsep\inst{6} \and
François Morain \inst{4}\fnmsep\inst{1} \and
Emmanuel Thomé \inst{1} \fnmsep\inst{2} \fnmsep\inst{3}
}

\institute{
Institut national de recherche en informatique et en
automatique (INRIA), \\ Villers-lès-Nancy \& Saclay, France \\
\and
Université de Lorraine, Loria, UMR 7503, Vandoeuvre-lès-Nancy, France
\and
CNRS, Loria, UMR 7503, Vandoeuvre-lès-Nancy, France 
\and
École Polytechnique/LIX, CNRS UMR 7161, Palaiseau, France
 \and
University of Calgary, Alberta, Canada
\and
Pacific Institute for the Mathematical Sciences, CNRS UMI 3069, Canada \\
\email{emmanuel.thome@inria.fr} \\
\email{morain@lix.polytechnique.fr} \\
\email{aurore.guillevic@ucalgary.ca} 
}

\makeatletter
\newsavebox\mytodobox
{%
\@tempdima\columnwidth\advance\@tempdima-2\fboxrule\advance\@tempdima-2\fboxsep
\begin{lrbox}{\mytodobox}\begin{minipage}{\@tempdima}\textbf{TODO:}}
  {\end{minipage}\end{lrbox}\par\noindent
  \ifdraft
  \fbox{\usebox{\mytodobox}}%
  \fi
  }
\makeatother

\let\maybeboldmu\relax
\let\maybeboldy\relax

\begin{document}

\maketitle

\begin{abstract}
Pairing based cryptography is in a dangerous position following the
breakthroughs on discrete logarithms computations in finite fields of
small characteristic. Remaining instances are built over finite fields
of large characteristic and their security relies on the fact the
embedding field of the underlying curve is relatively large. How
large is debatable. The aim of our work is to sustain the claim that the
combination of degree 3 embedding and too small finite fields
obviously does not provide enough security.
As a computational example, we solve the DLP on a 170-bit
MNT curve, by exploiting the pairing embedding to a 508-bit, degree-$3$
extension of the base field.
\end{abstract}

\section{Introduction}

Pairings were introduced as a constructive cryptographic tool in 2000 by
Joux~\cite{ANTS:Joux00}, who proposed a one-round three
participants key-exchange. Numerous protocols also
based on pairings have been
developed since. Beyond efficient broadcast protocols,
prominent applications include Identity-Based
Encryption~\cite{IEICE:KOS99,KOS00,C:BonFra01}, or short
signatures~\cite{AC:BonLynSha01}.

The choice of appropriate curves and pairing definitions in the context
of pairing-based cryptography has been the topic of many research
articles. An important invariant is the degree of the embedding
field, which measures the complexity of evaluating pairings, but is
also related to the security of systems (see Section
\ref{sec:background} for more precisions).
The first cryptographic setups proposed used pairings on
supersingular curves of embedding degree $2$ defined over
a prime field $\F_p$, where $p$ is 512-bit long, so that the pairing
embeds into a 1024-bit finite field $\F_{p^2}$. Another early curve
choice is a supersingular
elliptic curve in characteristic three, defined over $\F_{3^{97}}$, of
embedding degree $6$ (used e.g.\ in~\cite{AC:BonLynSha01}, as well as
various implementation proposals,
e.g.\,\cite{DBLP:journals/tc/BeuchatBDOST08}). More recent proposals
define \emph{pairing-friendly} ordinary curves over large
characteristic fields, where constraining the embedding degree to
selected values is a desired
property~\cite{ICISC:MiyNakTak00,CocksPinch01,JC:DupEngMor05,BreWen05,GalMcKVal07,ANTS:Freeman06,SAC:BarNae05,JC:FreScoTes10}.

Cryptanalysis of pairings can be attempted via two distinct routes. Either
attack the discrete logarithm problem \emph{on the curve}, or in the
embedding field of the pairing considered. The former approach is rarely
successful, given that it is usually easy to choose curves which are large
enough to thwart $O(\sqrt N)$ attacks such as parallel collision search
or Pollard rho. Note however that derived problems
such as the discrete logarithm \emph{with auxiliary
inputs} are much easier to handle, as shown by~\cite{PKC:SHITY12}.

Attacking pairings via the embedding field is a strategy
known as the \MOV~\cite{MenOkaVan93} or
\FR~\cite{FreRuc94} attack, depending on which pairing is considered.
Successful cryptanalyses that follow this strategy have been described in small characteristic. In~\cite{AC:HSST12},
for a supersingular curve over $\F_{3^{97}}$,
the small characteristic allowed the use of the
Function Field Sieve algorithm~\cite{AdHu99}, and the composite extension
degree was also a very useful property. More recently, following recent
breakthroughs for discrete logarithm computation in small
characteristic finite fields~\cite{EC:BGJT14,C:GraKleZum14}, a successful
attack has been reported on a supersingular curve over $\F_{2^{1223}}$,
with degree-$4$ embedding~\cite{C:GraKleZum14}. The outcome of
these more recent works is that curves in
small characteristic are now definitively avoided for pairing-based
cryptography.

As far as we know, there is no major record computation of discrete
logarithms over pairing-friendly curves in large characteristic using
a pairing reduction in the finite field. 
The pairing-friendly curves used in practice have a large embedding
field of more than 1024 bits, where computing a discrete logarithm 
is still very challenging. 
A few curves in large characteristic have comparatively small embedding
fields, and were identified as weak to
this regard, although no practical computation to date demonstrated the
criticality of this weakness.
This
includes the so-called MNT curves defined by \MNTnames, e.g.~\cite[Example 1]{ICISC:MiyNakTak00}, 
an elliptic curve defined over a
170-bit prime $p$, and of 508-bit embedding field $\F_{p^3}$.

Despite the academic agreement on the fact that the pairing embedding
fields for 170-bit MNT curves in general, and the one just mentioned in
particular, are too small for cryptographic use, recent work
like~\cite{CCS:ABDGGH15} has shown how cryptography relying on overly
optimistic hardness assumptions can linger almost indefinitely in the
wild. Demonstrating a practical break is key to really phasing out such
outdated cryptographic choices.
As far as we know, an MNT curve of low embedding degree $3$ was
never used in pairing-based cryptography, but was never attacked by a
pairing reduction either.
In this article, we present our attack over the weak\footnote{already
described as weak in the paper by the authors} MNT curve
\cite[Example~1]{ICISC:MiyNakTak00}, with $p$ of 170 bits and $n=3$. 
We report a
discrete logarithm computation in the group of points of this curve by a pairing
reduction, using only a moderate amount of computing power.

In order to attack the discrete logarithm problem in the embedding field,
appropriate variants of the Number Field Sieve must be used. The crucial
point is the adequate choice of a polynomial pair defining the Number
Field Sieve setup, among the various choices proposed in the
literature~\cite{C:JLSV06,Mat06,PAIRING:JouPie13,EC:BGGM15,AC:BarGauKle15}. It is also important to arrange
for the computation to take advantage of Galois automorphisms when
available, both within sieving and linear algebra. Last, some care is
needed in order to efficiently compute individual logarithms of arbitrary
field elements.

This article is organized as follows. Section~\ref{sec:background}
reviews some background and notations for MNT curves on the one hand, and
the Number Field Sieve (NFS) as a general framework on the other hand.
Section~\ref{sec:polyselect} discusses in more detail the various
possible choices of polynomial selection techniques for NFS.
Section~\ref{sec:dlp3} discusses the details of the discrete logarithm
computation with NFS, while Section~\ref{sec:challenge} defines and
solves an arbitrary challenge on the MNT curve.

\section{Background and notations}
\label{sec:background}

\subsection{Using pairing embedding to break DLP}
\label{sec:using-embedding}

We follow \cite[chap. IX]{BlSeSm05}.
To fix notations, 
pairings are defined as follows, the map being
bilinear, non-degenerate and
computable in polynomial time in the size of the inputs.
\begin{equation}\label{eqe}
    e:\left\{
\begin{array}{rccl}
     & E(\F_p)[\ell] \times  E(\F_{p^n})[\ell] & \to & \maybeboldmu{\mu}_\ell \subset \F_{p^n}^{*}\\
     & (P , Q) & \mapsto & e(P,Q).
\end{array}\right.
\end{equation}
Here, $\maybeboldmu{\mu}_\ell$ is the subgroup of $\ell$-th roots of
unity, i.e. an element $u \in \maybeboldmu{\mu}_\ell$ satisfies $u^\ell
= 1 \in \F_{p^n}^{*}$. The integer $n$ is the so-called {\em embedding
degree}, that is the smallest integer $i$ for which the $\ell$-torsion
is contained in $\F_{p^i}$. It has a major impact on evaluating the
difficulty of solving the DLP on the curve.

Let $G_1$ be a generator of $E(\Fp)[\ell]$ and $P$ in the
same group, whose discrete logarithm $u$ is sought (so that $P = [u]
G_1$). We choose
a generator $G_2$ for
$E(\F_{p^n})[\ell]$. We observe that
$$e(P,G_2)=e(G_1,G_2)^u$$
so that $u$ can be recovered as the logarithm of $U=e(P,G_2)$ in base $T
= e(G_1,G_2)$, where both
elements belong to the subgroup of order $\ell$ of $\FF_{p^n}^*$.
Note that by construction, $\ell = O(p)$, so that the Number Field
Sieve linear algebra phase has to be considered modulo $\ell$, which is
a priori much smaller than the largest prime order subgroup of
$\FF_{p^n}^*$, which has size $O(p^{\phi(n)})$.

\subsection{MNT curves}

The \MNTnames\ curves were designed in 2000 in~\cite{ICISC:MiyNakTak00} 
as the first example of
\emph{ordinary} curves with low embedding
degree $n=3$, $4$, or $6$.
The curves were presented as a weak instance of ordinary elliptic curves
that should be avoided in elliptic-curve cryptography because of the
\MOV\ and \FR\ attacks \cite{MenOkaVan93,FreRuc94} that embed the computation of a discrete
logarithm from the group of points of the curve to the embedding field
$\F_{p^n}$. At the 80-bit security level which was used in the 2000's,
an elliptic curve of 160-bit prime order was considered safe, and of
at least the same security as an 1024-bit RSA modulus. 
However for MNT curves over prime fields of 160 bits, the MOV and FR
reduction attacks embed to finite fields of size 480, 640, or 960 bits,
none of which should be considered as having a hard enough DLP.
For these three cases and most of all for $n=3$,
computing a discrete logarithm in the embedding field is considerably easier than over the
elliptic curve.
The conclusion of the MNT paper was to advise developers to
systematically check that the embedding degree of an elliptic curve is
large enough, to avoid pairing reduction attacks. 
The authors also mentioned as a constructive use of their curves the prequel work of  
Kasahara, Ohgishi, and Sakai on
identity-based encryption using pairings \cite{IEICE:KOS99,KOS00}.
Some implementations using MNT curves exist, for example the Miracl
Library proposes software on an MNT curve over a 170-bit prime, with
embedding degree $n=6$, providing a 80-bit security level.

\begin{table}
  \centering
  \begin{tabular}{|c|c|c|c|}
    \hline \begin{tabular}{@{}c@{}} embedding 
                                    degree $n$ \end{tabular}
                & $\log_2 p$ ($\# E(\Fp)$) & $n \log_2 p$ ($\# \F_{p^n}$) &
                                                    \begin{tabular}{@{}c@{}}
                                                      80-bit security\end{tabular} \\
    \hline 3 & 170 &  510 & no \\
    \hline 4 & 170 &  680 & no \\
    \hline 6 & 170 & 1020 & yes \\
    \hline
  \end{tabular}
  \caption{MNT curves as pairing-friendly curves in the 2000's}
  \label{tab:MNTcurves-size}
\end{table}

\subsubsection{Construction of MNT curves}
The parameters $p$, $\tau$, $\ell$ (base field, trace, and number of points) of
the curve are given by polynomials of degree at most two. For $n=3$,
$4$, or $6$, these are
$$\begin{array}{|c|l|l|l|}
    \hline \text{embedding degree }n &  p=P(x)   & \tau=\operatorname{Tr}(x)   & \#E(\Fp)
    = p+1-\tau \\
  \hline 3 & 12x^2 - 1 & \pm 6x - 1           & 12 x^2 \mp 6 x + 1 \\
  \hline 4 & x^2+x+1   & -x,\ \mbox{ or } x+1 & x^2 + 2 x + 2 \mbox{ or } x^2+1 \\
  \hline 6 & 4x^2 + 1  & 1 \pm 2x             & x^2 \mp 2 x + 1 \\
  \hline
\end{array}$$

To generate a curve, one needs to find an integer $\maybeboldy{y}$ of the
appropriate size, such that $p = P(\maybeboldy{y})$ is prime and $\#E(\Fp)$ is
also prime, or equal to a small cofactor times a large prime. 
To compute the coefficients of the curve equation, a Pell equation needs to be
solved. 

\subsubsection{The target curve}

Our target will be the MNT curve given in \cite[Example
1]{ICISC:MiyNakTak00}. We recall that the curve parameters satisfy 
$$\begin{array}{rcl}
\maybeboldy{y} &=& -\coeffs{0x732c8cf5f983038060466} \\
p &=& 12 \maybeboldy{y}^2 - 1 
   =  \coeffs{0x26dccacc5041939206cf2b7dec50950e3c9fa4827af} 
      \mbox{ of 170 bits}\\
    \tau &=& 6 \maybeboldy{y} - 1 \mbox{ where } \tau \mbox{ is the trace of the
  curve}\\
    \# E(\Fp) &=& p+1-\tau = 7^2 \cdot 313 \cdot \ell \mbox{ where } \ell \mbox{ is a
  156-bit prime}\\
\ell &=& \coeffs{0xa60fd646ad409b3312c3b23ba64e082ad7b354d} \\
\end{array}$$
The pairing embeds into the prime order $\ell$ subgroup of the cyclotomic
subgroup of $\FFpn{p}{3}$, where $\ell$ divides $p^2+p+1$.

\subsection{A brief overview of NFS-DL}

Our target field is $\FFpn{p}{n}$. NFS-DL starts by selecting two
irreducible integer polynomials~$f$ and~$g$
such that $\varphi=\gcd(f \bmod p, g\bmod p)$ is 
irreducible of degree $n$ (construction of~$f$ and~$g$ is
discussed in Section~\ref{sec:polyselect}).
We use the representation $\FFpn{p}{n} = \FF_p[x]/(\varphi(x))$. 
Let $K_f=\Q[x]/(f(x))=\Q(\alpha)$, and $\cO_f$ be
its ring of integers. Note that because $f$ is not necessarily monic,
$\alpha$ might not be an algebraic integer.
Let $\rho_f$
be the map from $K_f$ to $\FFpn{p}{n}$,
sending $\alpha$ to $T \bmod (p, \varphi(T))$.
We define likewise $K_g=\Q(\beta)$, together with $\cO_g$ and $\rho_g$.
This installs the (typical) commutative diagram in Figure~\ref{fig:diag}.

\begin{figure}
\begin{center}
\begin{tikzpicture}[>=latex]
    \node (Zx) at (5, 5) {$\Z[x]$};
    \node (Kf) at (3, 4) {$K_f$};
    \node (Kg) at (7, 4) {$K_g$};
    \node (Fpn) at (5, 3) {$\FFpn{p}{n} = \FF_p[x]/(\varphi(x))$};
    \draw[->] (Zx) -- (Kf);
    \draw[->] (Zx) -- (Kg);
    \draw[->] (Kf) -- node[yshift=-1ex,xshift=-0.25em,left]  {$\rho_f$} (Fpn);
    \draw[->] (Kg) -- node[yshift=-1ex,xshift=0.25em,right]  {$\rho_g$}(Fpn);
\end{tikzpicture}
    \caption{\label{fig:diag}NFS-DL diagram for \FFpn pn}
\end{center}
\end{figure}

Given $f$ and $g$, we choose a smoothness bound $B$ and build factor
bases $\mathcal{F}_f$ (resp. $\mathcal{F}_g$) consisting of prime
ideals in $\mathcal{O}_f$ (resp. $\mathcal{O}_g$) of norm less than
$B$, to which we add prime ideals dividing $\lc(f)$ (resp. $\lc(g)$) to take
into account the fact that $\alpha$ and $\beta$ are not algebraic
integers. Then, we collect relations, that is polynomials
$\phi(x) \in \Z[x]$ such that both ideals $\langle\phi(\alpha)\rangle$
and $\langle\phi(\beta)\rangle$ are \emph{smooth}, namely
factor completely over $\mathcal{F}_f$ (resp. $\mathcal{F}_g$).
Smoothness is related to $\Norm(\phi(\alpha))$, and in turn to $\Reslt(f,
\phi)$ since we have
$$\pm \lc(f)^{\mathrm{deg}(\phi)}  \Norm(\phi(\alpha)) = \Reslt(f, \phi).$$
When $\phi$ is such that the integers $\Reslt(f,
\phi)$ and $\Reslt(g,
\phi)$ are $B$-smooth (only prime factors below $B$), we have a relation:
$$\left\{\begin{array}{lcl}
\phi(\alpha) \mathcal{O}_f &=& \prod_{\mathfrak{q} \in \mathcal{F}_f}
\mathfrak{q}^{\mathrm{val}_{\mathfrak{q}}(\phi(\alpha))},\\
\phi(\beta) \mathcal{O}_g &=& \prod_{\mathfrak{r} \in \mathcal{F}_g}
\mathfrak{r}^{\mathrm{val}_{\mathfrak{r}}(\phi(\beta))}\\
\end{array}\right.$$
that are transformed as linear relation between virtual logarithms of
ideals~\cite{Schirokauer05}, to which are added the so-called Schirokauer
maps~\cite{Schirokauer93}, labelled
$\lambda_{f, i}$ for $1\leq i\leq r_f$ where $r_f$ is the unit rank of
$K_f$ (and the same for $g$).

To overcome the problem of dealing with fractional ideals instead of
integral ideals, we use the following result from
\cite{Montgomery97} (see also \cite{Huizing96b}). 
\begin{proposition}\label{prop:ideal-J}
    Let $f(X) = \sum_{i=0}^d c_i X^i$ with coprime integer
coefficients and $\alpha$ a root of $f$. Let
$$J_f = \langle c_d, c_d \alpha + c_{d-1}, c_d \alpha ^2+c_{d-1} \alpha
+ c_{d-2}, \ldots, c_d \alpha ^{d-1} + c_{d-1} \alpha ^{d-2} + \cdots
+ c_1\rangle.$$
Then $\langle 1, \alpha \rangle J_f = (1)$, $J_f$ has norm $|c_d|$, and
    $J_f \langle a-b \alpha\rangle$ is an integral ideal for integers $a$
    and $b$.
\end{proposition}
If $\phi(X)$ has degree
$k-1$, we have
$\Norm(J_f^{k-1} \langle\phi(\alpha)\rangle) = \pm \Reslt(f, \phi),$
so that we can read off the factorization of the integral $J_f^{k-1}
\langle\phi(\alpha)\rangle$ directly from the factorization of its
norm. A relation can now be written as:
$$(k-1) \vlog(J_f) + \sum_{\mathfrak{q} \in
\mathcal{F}_f}\mathrm{val}_{\mathfrak{q}}(\phi(\alpha))
\vlog(\mathfrak{q}) + \sum_{i=1}^{r_f} \lambda_{f, i}(\phi(\alpha))
\vlog(\lambda_{f, i})$$
$$\equiv (k-1) \vlog(J_g) + \sum_{\mathfrak{r} \in
\mathcal{F}_g}\mathrm{val}_{\mathfrak{r}}(\phi(\beta))
\vlog(\mathfrak{r}) + \sum_{i=1}^{r_g} \lambda_{g, i}(\phi(\beta))
\vlog(\lambda_{g, i})
\bmod \ell.$$

We select as many $\phi(x)$ of degree
at most $k-1$ (for $k \geq 2$ and very often $k=2$) as needed to find $\#
\mathcal{F}_f + \# \mathcal{F}_g + r_f + r_g + 2$ relations.
Note that $J_f$ and $J_g$ are not always prime ideals. Nevertheless since
all their prime divisors have a grouped contribution for each relation,
we may count them as single columns. We may even replace the two columns
by one, corresponding to $\vlog(J_f)- \vlog(J_g)$ (e.g. this is done in
\cadonfs).

Given sufficiently many equations, the linear system in the virtual logarithms
can be solved using sparse linear algebra techniques such as the
Block Wiedemann algorithm \cite{Coppersmith94}.
When we want to compute the logarithm of a given target, we need
to rewrite some power (or some multiple) of the target as a
multiplicative combination of the images in $\FFpn pn$ of the factor base
ideals, and use the precomputed data base of
computed logarithms.
Section~\ref{sec:dlp3} will briefly discusses 
algebraic factorization in practice.

\section{Polynomial Selection}
\label{sec:polyselect}
The polynomial selection is the first step of the NFS algorithm.
Polynomial selection is rather cheap, but care is needed since the
quality of the polynomial pair it outputs conditions the 
running time of the three next steps. 
Section~\ref{subsec:polyselect-method}
below explains the two phases of
polynomial selection. In a nutshell, we first decide from which family
the polynomials are chosen, and then we search among 
possible solutions for ``exceptionally good'' polynomials.
Note that because all degree $n$ irreducible polynomials correspond to
isomorphic finite fields $\FFpn pn$, we are not constrained in the choice
of $\Res(f,g)$. This degree of
freedom allows to select good polynomials.

As of 2016, the available polynomial selection algorithms are:
\begin{itemize}
\item the Conjugation method (Conj) \cite[\S.~3.3]{EC:BGGM15},
  explained in Algorithm~\ref{alg:Conj};
\item the Generalized Joux--Lercier method (GJL)
  \cite[\S.~3.2]{EC:BGGM15} and \cite{Mat06} that produces polynomials
  of unbalanced coefficient sizes; 
\item the Joux--Lercier--Smart--Vercauteren method (\JLSVi) \cite[\S.~2.3]{C:JLSV06},
  that produces two polynomials of degree $n$ and coefficient size in
  $O(\sqrt{p})$ for both polynomials;
\item the second proposition (\JLSVii) of the same paper \cite[\S.~3.2]{C:JLSV06};
\item the Joux--Pierrot (JP) method for pairing-friendly curves
  \cite{PAIRING:JouPie13} which
  produces polynomials equivalent to the Conjugation method for MNT
  curves;
\item the Tower-NFS method (TNFS) of Barbulescu, Gaudry and Kleinjung
  \cite{AC:BarGauKle15};
\item the Sarkar--Singh method that combines and generalizes the GJL
  and Conjugation methods \cite{EC:SarSin16}.
\end{itemize}

\begin{remark}[{Non-applicable methods.}]
The Extended-TNFS method of Kim and Kim-Barbulescu and its numerous variants
\cite{EPRINT:Kim15a,C:KimBar16,EPRINT:SarSin16:401,AC:SarSin16,EPRINT:SarSin16:537,EPRINT:JeoKim16:526} 
do not apply to finite fields of prime extension degree $n$ such as
$\F_{p^3}$. 
The TNFS method is not better than the best above methods for our
practical case study, as shown in the paper \cite[\S
5]{AC:BarGauKle15}. 
The Sarkar--Singh method \cite{EC:SarSin16} has two parameters
$(d,r)$: $d$ is a divisor of $n$ and $r\geq n/d$. Since
$n$ is prime, the pair $(d=1, r\geq n)$ corresponds to the GJL method
and the pair $(d=n,r=1)$ to the Conjugation method. 
The pair $(d=n,r=2)$ produces a polynomial
$f$ of degree 9 and small coefficients, and a polynomial $g$ of degree
6 and coefficients in $O(p^{1/3})$. This is not competitive for our
size of parameters $n=3$ and $p$ of 170 bits: the cross-over point
between the Conjugation ($r=1$) and their method ($r=2$) is at $\log_2
p^3 = 9592$ bits.
\end{remark}

Algorithm~\ref{alg:Conj} presents the Conjugation method, which eventually provided the best
yield. Pseudo-code describing the other methods can be found in
Appendix~\ref{sec:polyselect-appendix}.

\begin{algorithm}[hbtp]
\DontPrintSemicolon
\caption{Polynomial selection with the Conjugation method \cite[\S 3.3]{EC:BGGM15}}
\label{alg:Conj}
\KwIn{ $p$ prime and $n$ integer}
\KwOut{ $f,g,\psi$ with $f,g\in\Z[x]$ irreducible and $\psi=\gcd(f
\bmod p,g \bmod p)$ in $\FF_p[x]$ irreducible of degree~$n$
}
\Repeat {$a(y)$ has a root $\maybeboldy{y}$ in $\FF_p$ and
$\psi(x)=g_0(x)+ \maybeboldy{y} g_1(x)$ is irreducible in $\FF_p[x]$}
{
  Select $g_1(x), g_0(x)$, two polynomials with small integer coefficients,
  $\deg g_1 < \deg g_0 = n$ \;
  Select $a(y)$ a quadratic, monic, irreducible polynomial over
  $\Z$ with small coefficients \;
}
$f\gets \Reslt_y(a(y), g_0(x) + yg_1(x))$ \;
$(u,v)\gets$ a rational reconstruction of $\maybeboldy{y}$ \;
$g\gets vg_0 +u g_1 $ \;
\Return {$(f, g, \psi)$}
\end{algorithm}

\subsection{A First Comparison}
\label{subsec:polyselect-method}

\begin{figure}[htb]
  \centering
  \includegraphics[width=\textwidth]{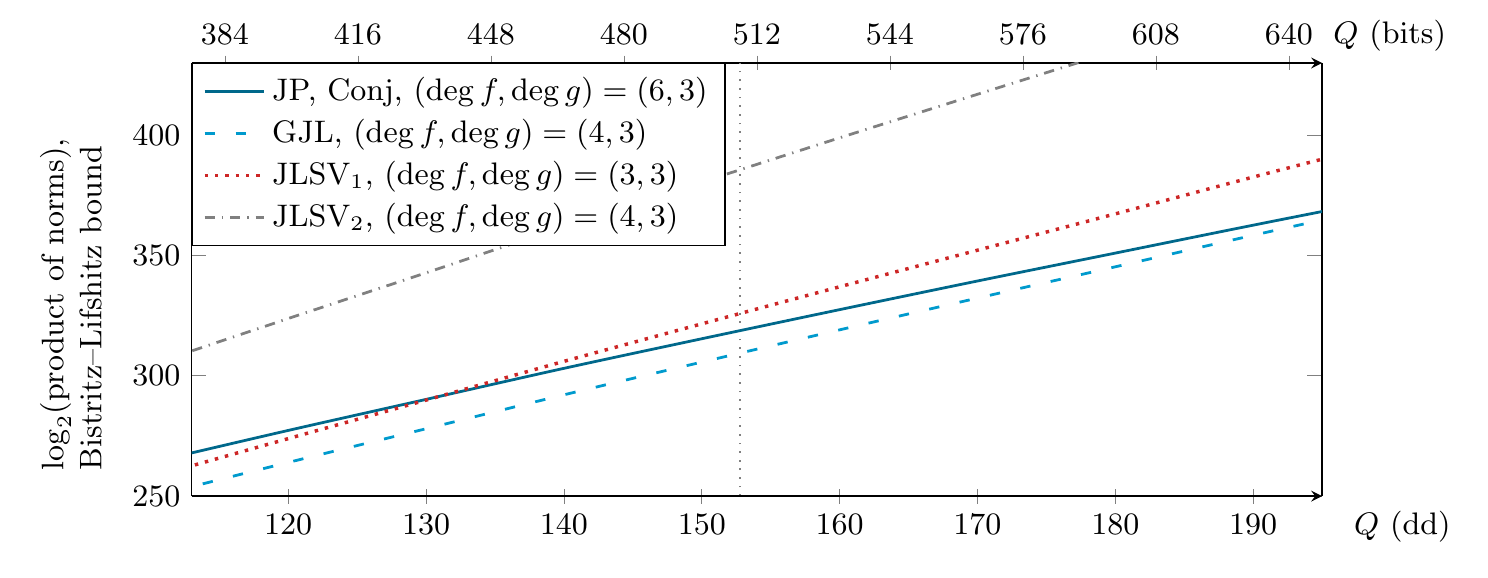} 
    \vskip -2\baselineskip
  \caption{Norm bound for four polynomial selection methods for $\F_{p^3}$}
  \label{fig:norm-bounds}
\end{figure}

The various methods above yield polynomial pairs whose characteristics
differ significantly. Table~\ref{tab:norm-bound} gives the expected
degrees and coefficient sizes. From this data, we can derive bounds on
the resultants on both sides of a relation (either using the coarse bound
$(\deg f + \deg \phi)!{\normoo f\!\!\!}^{\deg \phi}{\normoo \phi\!\!\!}^{\deg f}$, or
finer bounds such as~\cite[Th.~7]{ILAS:BisLif10}, as used
in~\cite[\S.~3.2]{AC:BarGauKle15}). These norms should be minimized in
order to obtain the best running-time for the NFS algorithm.  We obtain
the plot of Figure~\ref{fig:norm-bounds} for the bit-size of the product
of norms, similar to~\cite[Fig.~3]{EC:BGGM15}.

\begin{table}[htb]
  \centering
  \begin{tabular}{|c||c|c||c|c|}
\hline method   & $\deg f$   & $\| f \|_\infty$ & $\deg g$   & $\| g \|_\infty$ \\
\hline GJL      & $D+1 \geq n+1$ & $O(\log p)$ & $D \geq n$ & $O(Q^{1/(D+1)})$ \\
\hline JP or Conj & $2n$       & $O(\log p)$     & $n$        & $O(Q^{1/(2n)})$ \\
\hline \JLSVi & $n$        & $O(Q^{1/(2n)})$   & $n$        & $O(Q^{1/(2n)})$ \\
\hline \JLSVii & $D \geq n+1$ & $O(Q^{1/(D+1)})$ & $n$       & $O(Q^{1/(D+1)})$ \\
\hline
  \end{tabular}
  \caption{Norm bound w.r.t. $Q$}
  \label{tab:norm-bound}
\end{table}

Figure~\ref{fig:norm-bounds} suggests that 
the GJL method yields the smallest norms for $\log_2 Q = 508$. The norms
produced with the Conjugation and \JLSVi\ methods are not very far
however so we compared more precisely these three
methods for our 170-bit parameters. This entails finding competitive
polynomial pairs for each method, and comparing their merits.
Estimated bounds as well as experimental values for the products of norms 
for $\log_2 Q = 508$ are reported in Table~\ref{tab:norm-bound-generic}.
Results of sieving on one slide of special-$q$ is reported in
Table~\ref{tab:dry-runs}. 
The algorithms and computed
polynomials are given in Appendix~\ref{sec:polyselect-appendix}.
The theoretical bound $\|f\|_\infty$ equals one bit in the
Conjugation and GJL methods whereas in practice to improve the
smoothness properties of $f$, we have chosen a polynomial with moderately
larger coefficients, and with better
$\alpha$ and Murphy's $\EE$ values (see \cite[\S 5.2
eq.~(5.7)]{MurphyPhD99} on Murphy's $\EE$ value).
The coefficient size of $g$ selected with the GJL, Conj and \JLSVi
methods is a few bits
larger than the theoretical bound because we computed linear
combinations of two distinct $g$, and of $f$ and the initial $g$ in
the \JLSVi\ case (since they are of same degree).
The advantage of the hybrid Joux--Pierrot method
(Algorithm~\ref{alg:resultant}) in the MNT case is that 
$g$ can be monic, which does not allow for linear combinations. 

\begin{table}[htb]
  \centering
  \begin{tabular}{|l||r|r||r|r||c|c||c|c||c|c|}
\hline method   & \mc{2}{c||}{$\| f \|_\infty$} & \mc{2}{c||}{$\| g
  \|_\infty$} & \mc{2}{c||}{Norm bound $f$} & \mc{2}{c||}{Norm bound
  $g$} & \mc{2}{c|}{product} \\
\hline               & bound & exp. & bound & exp. & bound & exp. & bound
      & exp.  & bound & exp. \\
\hline GJL           &   1 &   2 & 127 & 130 & 106 & 107 & 206 & 208 & 311 & 314 \\
\hline Conj          &   1 &   9 &  85 &  86 & 157 & 165 & 163 & 164 & 320 & 328 \\
\hline hybrid JP     &   1 &  12 &  85 &  85 & 157 & 168 & 163 & 164 & 320 & 331 \\
\hline \JLSVi      &  85 &  85 &  85 &  86 & 163 & 163 & 163 & 164 & 326 & 327 \\
\hline \JLSVii      & 102 &  -- & 102 &  -- & 206 & --  & 180 &  -- & 386 & --  \\
\hline
  \end{tabular}
  \caption{Norm bounds in bits for $\log Q = 508$ and  $\log E = 25.25$:
      estimates based on
    Table~\ref{tab:norm-bound}, compared to experimental values with our
      selected polynomials.}
  \label{tab:norm-bound-generic}
\end{table}

\subsubsection{Galois actions.}
\label{subsec:polyselect-MurphyE}
\label{subsec:polyselect-JLSV1}

For small extension degrees $n \in \{3,4,6\}$ there exist families
of polynomials producing number fields with cyclic Galois groups, and
an easy-to-compute automorphism~\cite[Prop.~1.2]{Foster11}.
%
Taking polynomials from these families yields a speed-up in the sieving
part as well as in the linear algebra part for
the \JLSVi\ and Conjugation methods.
We take $g = x^3 - y_0 x^2 -(y_0+3) x - 1$
for the Conjugation method, i.e. $g_0 = x^3 -3x-1$ and $g_1 = -x^2-x $ in
Algorithm~\ref{alg:Conj}. 
The Galois action is $\sigma(x) =  (-x-1)/x$ which is independent of
the parameter $y_0$. 
In that case, given the factorization for $\langle a-b \alpha\rangle$,
we can deduce that of 
$$\sigma(\langle a-b \alpha\rangle) = \langle a-b
\sigma(\alpha)\rangle = -\frac{1}{\alpha} (b - (-a-b) \alpha).$$
The same holds on the $f$ side.

\subsubsection{Forming a database of good polynomials $f$.}

For the Conjugation method (and similarly for the competing methods), the early
steps in Algorithm~\ref{alg:Conj}
can be tabulated in some way, depending only on the extension degree
$n$ (and for \JLSVi, also on the size of $p$, but not its value):
we can store a database of $f$'s with
good smoothness properties (low $\alpha$ and high Murphy's $\EE$
values). Actually we searched over $a(y) = a_2y^2 + a_1y+a_0$,
where $0< a_2 < 32$, $|a_1| < 32$ and $|a_0| < 512$, and computed $f =
\Res_y(a(y), x^3 - yx^2 -(y+3)x - 1)$. Later, depending
on $p$, we can continue 
Algorithm~\ref{alg:Conj} for these precomputed polynomials (test
whether $a$ has a root modulo $p$).

Note also that in  Algorithm~\ref{alg:Conj}, the rational reconstruction
step naturally produces several quotients $u/v$, which yield several
candidate polynomials $g$. Small linear combinations of these polynomials
can be tried, in order to improve on the Murphy's $\EE$ value.

\subsection{Probing the sieving yield}
To finalize the comparison between the polynomials, we
compared the relation
yield
for small special-$q$ ranges sampled over the
complete special-$q$ space.
Because the \JLSVi\ and Conjugation methods feature balanced
norms (see Tab.~\ref{tab:norm-bound-generic}), we used similar large
prime bounds (27 bits) on both sides in both cases, and 
allowed two large prime on each side. In contrast, for the GJL method,
we allowed 28-bit large primes on the $g$ side, and chose $q$
to be only on that side.
The Conjugation method (polynomial below)
appeared as the best option based on the seconds/relation measure, given
that the overall yield was sufficient.
Results of this test are reported on Table~\ref{tab:dry-runs}.

\label{sec:polys-used}
\begin{equation}
\label{eq:polys-used}
\begin{array}{l@{}cl}
f &=&
    \coeffs{28}x^6 + \coeffs{16}x^5 - \coeffs{261}x^4 - \coeffs{322}x^3 +
    \coeffs{79}x^2 + \coeffs{152}x + \coeffs{28}\\
\alpha(f) &=& -2.94 \\
\log_2 \|f\|_\infty &=& 8.33\\
g &=& 
    \coeffs{24757815186639197370442122}\ x^3 +
    \coeffs{40806897040253680471775183}\ x^2 \\&& -
    \coeffs{33466548519663911639551183}x -
    \coeffs{24757815186639197370442122}\\
\alpha(g) &=& -4.16 \\
\log_2 \|g\|_\infty &=& 85.08, \mbox{ the optimal being
}\frac{1}{2}\log_2 p = 85 \\
\EE(f,g) &=& 1.31 \cdot 10^{-12}
\end{array}
\end{equation}

\begin{table}
    \begin{center}
        \begin{tabular}{|l|c|c|l|}
            \hline
            Method & seconds/relation & relations/special-$q$ & remarks\\
            \hline
            \begin{tabular}{@{}l@{}} Generalized \\Joux--Lercier\end{tabular} &
 3.48 & 4.96
            & 0+3 large primes below $2^{28}$\\
            \hline
            \JLSVi &
 1.31 & 4.24
            &\multirow{2}{15em}{2+2 large primes below $2^{27}$, orbits of three
            special-$q$ batched together}\\
            \cline{1-3}
            Conjugation &
{\bfseries 0.91} & 5.93
            &\\\hline
        \end{tabular}
            \caption{\label{tab:dry-runs}Probed yield for special-$q$
            ranges.
            Cpu time on Intel Xeon E5520 (2.27GHz).}
    \end{center}
\end{table}

\section{Solving DLP over $\FFpn{p}{3}$}
\label{sec:dlp3}

\subsection{Sieving and linear algebra}

We took a smoothness bound of $50\times10^6$ on both sides; and all special-$q$
in $[50\times 10^6, 2^{27}]$, on both sides. This took roughly 660 core-days,
normalized on the most common hardware used, namely 4-core Intel Xeon
E5520 CPUs (2.27GHz).
We collected 57070251 relations, out of which 34740801 were non
duplicate.
Filtering produced a
$1982791\times 1982784$ matrix $M$ with weight
$396558692$. Taking into account the block of
7 Schirokauer maps $S$, the matrix $M\|S$ is square.

We computed 8 sequences in the Block Wiedemann algorithm,
using the trick mentioned in \cite[§8]{Coppersmith94}, as
programmed in \cadonfs\ (rediscovered and further analyzed in
\cite{JoPi15}). All these sequences can be computed independently. 
Computation time for the 8 Krylov sequence was about 250 core-days (Xeon
E5-2650, 2.4GHz, using four 16-core nodes per sequence).
Finding
the linear (matrix) generator for the matrices took 75 core-hours,
parallelized over 64 cores.
Building the solution cost some more 170 core-days. We
reconstructed virtual logarithms for 15196345 out of the
15206761
factor base elements (99.9\%).
This was good enough to start looking for
individual logarithms.

\subsection{Computing individual discrete logarithms in $\FFpn{p}{3}$}
\label{ssct:ind-log}

From the linear algebra step, we know how to compute the logarithm modulo
$\ell$ of any element of $\FFpn{p}{3}$ whose lift in either $K_f$
or $K_g$ factors completely over the factor base. Lifting in $K_f$ is
often convenient because norms are smaller.

\paragraph{The tiny case.}
A particular element which lifts conveniently in $K_f$ is the common root
$t$ of both polynomials. By construction, its lift $\alpha\in K_f$
generates a principal (fractional) ideal that factors as $J_f^{-1}$ (see
Proposition~\ref{prop:ideal-J}) times prime
ideals of norm dividing 28, namely:
$(\alpha)=I_{2,0}^2I_{2,\infty}^{-2}I_{7,0}I_{7,\infty}^{-1}$,
where $I_{2,\infty}^{2}I_{7,\infty}$ corresponds to $J_f$ and the prime ideals in the right-hand side can be made explicit.
Its logarithm therefore writes as
\footnote{The convention in \cadonfs\ is to take coefficients of
largest degree first in the Schirokauer maps computation
$z\mapsto \frac1\ell(z^{\ell^m-1}-1)$ where
$m=\operatorname{lcm}_{\mathfrak{l}\
\text{prime},\ \mathfrak{l}\mathrel|\ell}
[\mathfrak{l}:\ell]$. Here we have $m=1$.}
\begin{align*}
    \log(t)&=
        2\vlog{I_{2,0}}-2\vlog{I_{2,\infty}}+\vlog{I_{7,0}}-\vlog{I_{7,\infty}}+
    \sum_{i=1}^5\lambda_{f,i}(\alpha)\vlog(\lambda_{f,i}).\\
    \lambda_{f,1}(\alpha)&=\coeffs{0x3720106a3d368d7f731a0757b905778050ae327},
    \lambda_{f,2}(\alpha)=\coeffs{0x1dbeace7d0ec187712ae8afcd6ccdc4db06f781},\\
    \lambda_{f,3}(\alpha)&=\coeffs{0x9c3109f7741d625869f135706be03fc09375450},
    \lambda_{f,4}(\alpha)=\coeffs{0x1e46653b287d99c502a5c6e12ab17a3dd10988c},\\
    \lambda_{f,5}(\alpha)&=\coeffs{0x31628f3e0b491e622946b32f66292c1389a7427}.
\end{align*}
By construction the value $\log(t)$ above is invertible modulo $\ell$,
and we can freely normalize our virtual logarithm values so that it
is equal to one.

\paragraph{The tame case.}
Elements whose lifts do not factor completely over any of the factor
base but have only moderate-size outstanding factors can be dealt
with using a classical \emph{descent} procedure. This finds recursively
new relations involving smaller and
smaller primes, until all primes involved belong to the factor base.
Software achieving this exists, such as the  \verb+las_descent+
program in \cadonfs.

\paragraph{The general case.}
For computing individual logarithms of arbitrary elements, 
we used the boot technique described in~\cite{AC:Guillevic15}.
For each target, we compute a preimage in $\ZZ[x]$ represented
by a polynomial of degree at most 5 and coefficients bounded by
$p^{1/3}$. The norm in $K_f$ of the preimage
is $O(p^2) = O(Q^{2/3})$, of approximately 340 bits.
The asymptotic complexity of this step is $L_Q[1/3, 1.26]$, and would be
$L_Q[1/3, 1.132]$ with one early-abort test
(see e.g.~\cite[\S 4.3]{Pomerance82} or~\cite[Ch.~4]{BarbulescuPhD13}). The
optimal size of largest prime factors in the decomposition is given by
the formula $L_Q[2/3, (e^2/3)^{1/3} \approx 0.529]$, where $e = 2/3$
(see \cite[\S 4]{PKC:ComSem06}). Applying it for
$\log_2 Q = 508$ gives a bound of 68 bits and a running-time of
approximately $2^{42}$ tests.
In practice we found very easily initial splittings where $B_1$ is
less than 64 bits, 
which eased the descent.


\subsection{Solving the challenge}
\label{sct:challenge}
\label{sec:challenge}

Our main use case for individual
logarithm computation in \FFpn p3 is to solve a DLP challenge on the curve.
The challenge definition procedure (described in the appendix%
\footnote{\S B.1 and \S B.2 of the pre-proceedings version available at
\url{https://hal.inria.fr/hal-01320496}}, 
the Magma code is also available%
\footnote{\label{footnote:magma-script}\url{http://www.lix.polytechnique.fr/~guillevic/discrete-log/SAC2016-mnt170-verification-script.mag}})
gives:
\begin{align*}
    G_1 &= (\coeffs{0x106b415d7b4a2d71659ae97440cbb20a6de42d76d69},
    \coeffs{0x16d74a2a88e817f1821a1c40e220d34eec93e33cb83}),\\
    P&=(\coeffs{0x15052ba45717710e6b0cbf8ed89c5c1a0a279480e26},
\coeffs{0x8050f05a231ae1f13e56de1171c108294656052339})
\end{align*}

From Section~\ref{sec:using-embedding}, we need to compute $\log(G_T)$ and
$\log(S)$, where $G_T=e(G_1, G_2)$ and $S=e(P,G_2)$
are given in the Magma verification script%
\footnotemark[\getrefnumber{footnote:magma-script}]. We searched for
randomized values $G_T^r$ and $G_T^{r'} S$ which were amenable to the descent procedure.
After 32 core-hours looking in the range $r\in[1, 64000]$,
we selected the following element
\begin{align*}
    G_T^{52154} &=
 - \coeffs{0x21d517d51512e9} -\coeffs{0x95233b3af1b3c7}\, x + \coeffs{0x8d324ebc7849bb}\, x^2 \\
& + \coeffs{0x18ff0d5ae0b52b}\, x^3 + \coeffs{0x13f711fe92d63cd}\, x^4 -\coeffs{0x15c778630d36920}\, x^5
\end{align*}
whose straightforward lift in $K_f$ has 59-bit smooth norm (resultant
with $f$, more precisely):
\begin{gather*}
\coeffs{0x87ac1a057df9772d1e08d4de56b3e6b5f208710437b5f92ac4a494c318c9781107e00364934e34efa87b26597771c}\\
=\coeffs{2^2}
 \cdot \coeffs{5}
 \cdot \coeffs{7^2}
 \cdot \coeffs{31}
 \cdot \coeffs{193}
 \cdot \coeffs{277}
 \cdot \coeffs{1787}
 \cdot \coeffs{12917}
 \cdot \coeffs{125789}
 \cdot \coeffs{142301513}
 \cdot \coeffs{380646221}
 \cdot \coeffs{2256567883}\\
 \cdot \coeffs{132643203397}
 \cdot \coeffs{138019432565816569}
 \cdot \coeffs{603094914193031251}
 \cdot \coeffs{801060739300538627}
\end{gather*}
Virtual logarithms for primes below $50\cdot 10^6$ (25.57 bits) were known.
 The descent procedure took
 13.4 hours. 
Once all logarithms were computed, the value of
$\log(G_T)$ could be deduced:
$$ \log(G_T) = \coeffs{0x8c58b66f0d8b2e99a1c0530b2649ec0c76501c3}
\text{ (normalized to $\log t=1$)}.$$
Similarly, we selected
\begin{align*}
G_T^{35313}S \mapsto &
\phantom{+}    \coeffs{0x457449569db669} + \coeffs{0x88c32ec54242fd}\, x - \coeffs{0x2370c0f5914ba9}\, x^2\\
& + \coeffs{0x14c7ccbafc20e2}\, x^3 + \coeffs{0xde2e21c5f1a4c4}\, x^4 - \coeffs{0x10b6bfd826db49c}\, x^5
\end{align*}
whose lift in $K_f$ has norm
\begin{gather*}
-\coeffs{0x44dafd6ec57c91e64567fa045187100da9a98c5c509b388cb61759f345b3ce27226a5e8520be0bd4559acbd538b90}\\
=-\coeffs{2^4}
\cdot \coeffs{5^2}
\cdot \coeffs{7}
\cdot \coeffs{643}
\cdot \coeffs{1483}
\cdot \coeffs{2693}
\cdot \coeffs{95617}
\cdot \coeffs{9573331}
\cdot \coeffs{33281579}
\cdot \coeffs{1608560119}
\cdot \coeffs{48867401441}\\
\cdot \coeffs{516931716361}
\cdot \coeffs{896237937459937}
\cdot \coeffs{16606283628226811}
\cdot \coeffs{19530910835315983}
\end{gather*}
the largest factor having 54 bits, a very small size indeed (compared
to the 68 bits predicted by theory). 
The descent
procedure for other primes took
10.7 hours.
We found that
$$ \log(S) = \coeffs{0x48a6bcf57cacca997658c98a0c196c25116a0aa}
\text{ (normalized to $\log t=1$)}.$$

We  eventually 
found that 
$$\log_{G_1} (P) = \coeffs{0x711d13ed75e05cc2ab2c9ec2c910a98288ec038} \bmod \ell.$$

\section{Conclusion and future work}

\subsection{Consequences for pairing-based cryptography}
Our work showed that the choice of embedding degree $n$ and finite field
size $\log p^n$ should be done carefully. The size of $\F_{p^n}$ should be large
enough to provide the desired level of security. We recall these sizes
for $\F_{p^3}$.
The recent improvements of Kim and Kim--Barbulescu
\cite{EPRINT:Kim15a,C:KimBar16} do not apply to  
$\F_{p^n}$ where $n$ is prime, so $\F_{p^3}$ is not affected. 
The asymptotic complexity of the NFS algorithm for $\F_{p^3}$ is $ \exp\left(
  (c+o(1)) (\log p^n)^{1/3} (\log \log p^n)^{2/3} \right) = L_{p^3}[1/3,
(64/9)^{1/3}]$. Since there is a polynomial factor hidden in the
notation $c+o(1)$, taking $\log_2 L_{p^3}[1/3, (64/9)^{1/3}]$ does not
give the exact security level but only an approximation.
We may compare our present record with previous records of same size
for prime fields $\F_p$ and quadratic fields $\F_{p^2}$.
Kleinjung in 2007 announced a record computation in a prime field
$\F_p$ of 530 bits (160 decimal digits) \cite{NMBRTHY:Kleinjung07}.
Barbulescu, Gaudry, Guillevic and Morain in 2014 announced a record
computation in $\F_{p^2}$ of 529 bits (160 decimal digits) \cite{NMBRTHY:BGGM14}.
We compare the timings in Table~\ref{tab:timings-DL-records}.
The timings of relation collection and linear algebra were not
balanced in Kleinjung record: 3.3 years compared to 14 years and
moreover, this is a quite old record so it is not really possible to
compare our record with this one directly. 
We can compare our record with the 529-bit $\F_{p^2}$ record
computation of 2014 \cite{NMBRTHY:BGGM14}. Our total running-time is
15.5 times longer whereas the finite field is 21 bit smaller.

\begin{table}
  \centering
  \begin{tabular}{|l|c|c|c||c|}
\hline                               record & relation collection & linear algebra & individual log  & total      \\
\hline Kleinjung \cite{NMBRTHY:Kleinjung07} & 3.3 CPU-years       & 14 years       & few hours       &            \\
       530-bit field                        & 3.2 GHz Xeon64      & 3.2 GHz Xeon64 & 3.2 GHz Xeon64  & 17.3 years \\
          $\F_p$, 2007                      &                     &                &                 &            \\
\hline BGGM \cite{NMBRTHY:BGGM14}           & 68 core-days=0.19y  & 30.3 hours     & few hours       &  70 days   \\
       529-bit field                        &    2.0 GHz E5-2650  & NVidia GTX 680 & 2.0 GHz E5-2650 & = 0.2 year \\
          $\F_{p^2}$, 2014                   &                     &   graphic card &                 &            \\
\hline BGGM \cite{CATREL:BGGM15}            & 850 core-days       & 5500 core-days & few days        &            \\
       512-bit field                        &  = 2.33 years       & = 15 years     &                 & 17.3 years \\
\cline{2-4} $\F_{p^3}$, 2015                 &\mc{3}{c||}{2.4 GHz Xeon E5-2650}&\\
\hline this work                            & 660 core-days       & 423 days        & 2 days         & 1085 days  \\
       508-bit field                        & =1.81 years         &  = 1.16 years   &                & = 2.97 years\\
          $\F_{p^3}$, 2016                   & 2.27GHz 4-core      & 2.4 GHz         & 2.27GHz 4-core &            \\
                                            &  Xeon E5520         & Xeon E5-2650    &  Xeon E5520    &            \\
\hline
  \end{tabular}
  \caption{Comparison of running-time for Discrete Logarithm records
    in $\F_p$, $\F_{p^2}$ and $\F_{p^3}$ of 530, 529, 512 and
  508 bits.}
  \label{tab:timings-DL-records}
\end{table}

\subsection{Future work}

We have computed a DLP on an MNT curve with embedding degree 3. What
are the next candidates?
We could continue the series in two directions: increasing the size of
$p^n$ to 600 bits, in order to compare this new record to the
previous records of the same size, in particular the $\F_{p^2}$
record of 600 bits \cite{EC:BGGM15}. We could conjecture, according to
the present record and the size of the norms, that a DLP record in $\F_{p^3}$ of 600 bits
will be more than 15 times harder than in a 600-bit field $\F_{p^2}$.

The second direction would be to continue the series of MNT curves,
with $n=4$. We found an MNT curve of embedding degree 4 in Miracl
(file \texttt{k4mnt.ecs}). The 
curve was generated by Drew Sutherland for Mike Scott a long time ago.

$$\begin{array}{rcl}
y &=& \coeffs{0xf19192168b16c1315d33} \\
p &=& y^2+y+1 = \coeffs{0xe3f367d542c82027f33dc5f3245769e676a5755d} \\
\ell &=& \coeffs{0x6b455e0a014f1e30eaef7300bd4bb4258290fc5} \\
\tau &=& y+1 =\coeffs{0xf19192168b16c1315d34}\\
\# E(\Fp) &=& y^2 + 1 = p+1-\tau = 2\cdot 17 \cdot \ell \\
\end{array}
$$

Since $n$ is a prime power, we have to adapt the
Kim--Barbulescu technique (dedicated to non-prime power $n$) to
prime-power extension degrees\footnote{right after the submission,
  several variants of Kim's Extended TNFS where proposed, that
  deal with any composite $n$, in particular prime power $n$, and
  generalize the Sarkar--Singh method
  \cite{EPRINT:SarSin16:401,AC:SarSin16,EPRINT:SarSin16:537,EPRINT:JeoKim16:526}.}. 
We construct $\F_{p^4}$ as $\F_{p^2}[x]/(\varphi(x))$, where
$\F_{p^2} = \F_p[s]/(h_1(s))$ and both $h_1$ and $\varphi$ are of degree
2, and $\varphi$ has coefficients in $\F_{p^2}$. As a consequence,
the polynomials $f$ and $g$ will have coefficients in
$\ZZ[s]/(h_1(s))$ instead of $\ZZ$. For example, one could take
$$\begin{array}{rcl}
h_1(s) &=& s^2 + 2, \\
h_2(x, t_0, s) &=& x^2 + s + t_0, \\
P(t_0) &=& t_0^2+t_0+1, \\
f &=& {\Reslt}_{t_0}(P(t_0), h_2(x, t_0, s)) = x^4 + (2s - 1)x^2 - s - 1,\\
g &=& h_2(x, y, s) = x^2 + s + \coeffs{0xf19192168b16c1315d33}.
\end{array}$$
The major difference is that to be efficient, we have to sieve
polynomials of degree 1 with coefficients in $\ZZ[s]/(h_1(s))$, that is
elements of the form $(a_0 + a_1 s) + (b_0 + b_1 s) x$ where the
$a_i$'s and $b_i$'s are small rational integers, say $|a_i|, |b_i|
\leq A$. For instance,
taking $\log _2 (E) = 1.1 (\log Q)^{1/3} (\log \log Q)^{2/3} \approx
28$, we obtain $A = E^{2/(2 \deg h)}$ of 14 bits.
The upper bound on the norm would be of 120 bits on $f$-side and 219
bits on $g$-side, the total being roughly of 339 bits. This is 11 bits
more than our present record for the 508-bit $n=3$ MNT curve (328
bits, Table~\ref{tab:norm-bound-generic}), but by
far much less than with any previous technique applied to that
$\F_{p^4}$. Norm estimates are
provided in Table~\ref{tab:MNT4-norm-bounds}. From a practical point
of view, we would need extensions of the work \cite{ANTS:GauGreVid16}.

\begin{table}[htpb]
  \centering

  \caption{Norm bound estimates for $\F_{p^4}$ of 640 bits.}
\label{tab:MNT4-norm-bounds}
  \begin{tabular}{|l|r|r|c|c|c|}
\hline method   & $\| f \|_\infty$ & $\| g \|_\infty$ & 
$NB_f$& $NB_g$ & $NB_f + NB_g$ \\
\hline Extended TNFS+hybrid JP&   1 &  80 & 120 & 219 & 339 \\
\hline GJL                    &   1 & 128 & 144 & 243 & 387 \\
\hline \JLSVi                 &  80 &  80 & 195 & 195 & 390 \\
\hline Sarkar-Singh, $d=2,r=2$&   1 & 107 & 172 & 222 & 394 \\
\hline Hybrid JP--Conj        &   1 &  80 & 159 & 240 & 399 \\
\hline \JLSVii, $D=6$ ($D$ best choice) &  91 &  91 & 264 & 206 & 470 \\
\hline
  \end{tabular}
\end{table}

\paragraph{Acknowledgements.}
The authors are grateful to Pierrick Gaudry for his help in running
the computations.

\bibliographystyle{abbrv}

\appendix

\section{Polynomial selection methods}
\label{sec:polyselect-appendix}

We provide in this section the polynomials computed for our $\F_{p^3}$ record
with the other competitive polynomial selection
methods that we compared in Section~\ref{sec:polyselect}.

\paragraph{Generalized Joux--Lercier method.}

The first step of the GJL polynomial selection algorithm is to choose a polynomial
$f$ of degree 4 in our context. We need $f$ to factor as a linear
polynomial times a degree 3 polynomial modulo $p$, hence we cannot
allow for a degree two subfield, or any of the Galois groups C4, V4 or D4.
We extracted from the Magma number field database the list of irreducible
polynomials of degree 4 and Galois group A4 (of order 12), class
number one and signature $(0,2)$ (592 polynomials) and $(4,0)$ (3101 polynomials).

In the GJL method, the LLL algorithm outputs four
polynomials $g_1$, $g_2$, $g_3$ and $g_4$ with small coefficients. 
To obtain the smallest possible coefficients, we set the LLL
parameters to $\delta = 0.99999$ and $\eta = 0.50001$.
We compute
linear combinations $g = \sum_{i=1}^{4} \lambda_i g_i$ with
$|\lambda_i| \| g_i\|_\infty \leq 2^5 \cdot \min_{1\leq i\leq 4} \|g_i\|_\infty$
(roughly speaking, $|\lambda_i| \leq 32$)
so that the size of the coefficients of $g$ do not increase too much,
while we can obtain a polynomial $g$ with a better Murphy's $\mathbb{E}$ value. 

Then we run the GJL method with our modified
post-LLL step for each polynomial $f$ in our
database and we selected the pair with the highest Murphy's $\mathbb{E}$
value. We obtained 
$$\begin{array}{lcl}
f &=& x^4 - 2 x^3 + 2 x^2 + 4 x + 2 \\
\alpha(f) &=& 1.2 \\
\log_2 \|f\|_\infty &=& 2\\
\multicolumn{3}{l}{ g = \mathsf{\scriptstyle{133714102332614336563681181193704960555}}
     \hspace*{2pt} x^3
     + \mathsf{\scriptstyle{173818706907699496668994559342802299969}}\hspace*{2pt} x^2}\\
\multicolumn{3}{l}{\phantom{g=} + \mathsf{\scriptstyle{878019651910536420352249995702628405053}}\hspace*{2pt}  x
     - \mathsf{\scriptstyle{185403948115503498471378323785210605885}}} \\
\alpha(g) &=& -2.1 \\
\log_2 \|g\|_\infty &=& 129.37, \mbox{ the optimal being
}\frac{3}{4}\log_2 p = 127.5\\
\EE(f,g) &=& 5.08 \cdot 10^{-13} \\
\end{array}$$

\paragraph{Joux-Lercier-Smart-Vercauteren method.}

The Joux-Lercier-Smart-Vercauteren method (JLSV1)
is possibly the most straighforward
polynomial selection method adapted to non-prime finite fields. It is
possible to force this method to pick polynomials $f$ within a specific
family, in order to force nice Galois properties. For example, we may use
the form $\psi = x^3- tx^2 -(t+3)x-1$.

The enumeration was the largest for the \JLSVi\ method:
we searched over $2^{25}$ polynomials $f$ in the cyclic family 
$x^3 - t_0x^2 - (t_0+3)x-1$, with a parameter $t_0$ of 84 up to 85 bits. 
We kept the polynomials whose $\alpha$ value
was less than $-3.0$. We continued the \JLSVi\ polynomial selection
algorithm 
selectively for these good precomputed polynomials.
The ``initial'' $g$ (say $g_0$) produced by the method can
be improved by using instead any linear combination $g = \lambda f + \mu
g_0$ for small
$\lambda$ and $\mu$, thereby improving the Murphy's $\EE$ value.
We set $|\lambda|, |\mu| \leq 2^5$.

\begin{equation}
\begin{array}{l@{}cl}
f &=& x^3 - \mathsf{\scriptstyle{30145663100857939296343446}}\hspace*{2pt} x^2
 - \mathsf{\scriptstyle{30145663100857939296343449}}\hspace*{2pt} x - 1 \\
\alpha(f) &=& -3.0 \\
\log_2 \|f\|_\infty &=& 84.64\\
g &=& \mathsf{\scriptstyle{30145663100857939299699540}}\hspace*{2pt} x^3
     + \mathsf{\scriptstyle{46845274144495980578316407}}\hspace*{2pt} x^2\\
 & & - \mathsf{\scriptstyle{43591715158077837320782213}}\hspace*{2pt} x
     - \mathsf{\scriptstyle{30145663100857939299699540}} \\
\alpha(g) &=& -2.8 \\
\log_2 \|g\|_\infty &=& 85.28, \mbox{ the optimal being
}\frac{1}{2}\log_2 p = 85 \\
\EE(f,g) &=& 1.02 \cdot 10^{-12}
\end{array}
\end{equation}

\paragraph{Conjugation and Joux--Pierrot methods.}
The Joux-Pierrot method produces polynomials with the same degree and
coefficient properties
as the Conjugation method for MNT curves and that are moreover monic. 
The polynomials constructed with the Conjugation method allow a factor
two speed-up thanks to a Galois automorphism. 
We propose here a hybrid variant in Algorithm~\ref{alg:resultant}
for pairing-friendly curves. The conjugation method, in
Algorithm~\ref{alg:Conj},  is the one which eventually produced the best
polynomial pair.

For the Conjugation method as well as the hybrid method of
Algorithm~\ref{alg:resultant}, and similarly to the \JLSVi\ method, it is possible to choose polynomials $g$
of the form $\psi = x^3- tx^2 -(t+3)x-1$
to allow a Galois automorphism of degree 3.

\begin{algorithm}[hbtp]
  \DontPrintSemicolon
  \caption{Variant of Joux--Pierrot and
    Conjugation methods}
  \label{alg:resultant}
  \KwIn{ $p$ prime, $p = P(\maybeboldy{y})$ where $\deg P \geq 2$ and
    $P$ of tiny coefficients, $n$ integer}
  \KwOut{ $f,g\in\Z[x]$ irreducible and $\psi=\gcd(f
    \bmod p,g \bmod p)$ in $\FF_p[x]$ irreducible of degree~$n$
  }
  \Repeat {$\psi(x)$ is irreducible in $\FF_p[x]$ and $f$, $g$ are
    irreducible in $\ZZ[x]$}{
    Select $g_1(x), g_0(x)$, two polynomials with small integer coefficients,
    $\deg g_1 < \deg g_0 = n$ \;
    Select small integers $a,b,c,d$ \;
    $\psi(x) = g_0(x)+ \left( \dfrac{a+b\maybeboldy{y}}{c+d \maybeboldy{y}} \bmod p \right) g_1(x)$\;
  $f\gets \Reslt_Y\left(P(Y), (c+dY) g_0(x) + (a+bY)g_1(x)\right)$ \;
  $g\gets  (c+d \maybeboldy{y}) g_0(x) + (a+b \maybeboldy{y}) g_1(x)$
  \tcp*[r]{$g \equiv  (c+d \maybeboldy{y}) \psi(x) \bmod p$}
  }
  \Return {$(f, g, \psi)$}
\end{algorithm}

In practice, in Algorithm~\ref{alg:resultant} one might prefer to
constrain $d=0$, so that  $g$ has small leading
  coefficient $c$. Going further and requiring $c=1$ so that $g$ is
  monic reduces however too much
  the
  possibilities to find a good pair of polynomials.

  The following example has been obtained with
  Algorithm~\ref{alg:resultant}, 
searching over all
$(a + b \maybeboldy{y})/c$ with $|a|, |b|, |c| \leq 256$.
$$\begin{array}{rcl}
\maybeboldy{y} &=& -\coeffs{8702303353090049898316902} \mbox{ is the
  targeted MNT curve parameter}\\
f &=& 108 x^6 + 1116 x^5 + 3347 x^4 + 2194 x^3 - 613 x^2 - 468 x + 108 \\
g &=& 6 x^3 + \coeffs{34809213412360199593267639} \ x^2 +
\coeffs{34809213412360199593267621} \ x - 6 \\
  &=& 6 x^3 - (4 \maybeboldy{y} - 31)x^2 - (4 \maybeboldy{y} - 13)x - 6 \\
\varphi &=& \frac{1}{6} g \bmod p = x^3 +
\coeffs{151460167298404651346258165094598961506004769966481}\ x^2 \\
 & & + \coeffs{151460167298404651346258165094598961506004769966478} \ x - 1\\
\end{array}$$

\end{document}